\documentclass{webofc}
\usepackage[varg]{txfonts}   
\usepackage{paralist}
\usepackage{placeins}
\usepackage{xcolor}

\begin{document}

\title{Track Reconstruction in the ALICE TPC using GPUs for LHC Run~3}

\author{\firstname{David} \lastname{Rohr}\inst{1}\fnsep\thanks{\email{drohr@cern.ch}} \and
        \firstname{Sergey} \lastname{Gorbunov}\inst{2,3}\fnsep\thanks{\email{gorbunov@compeng.uni-frankfurt.de}} \and
        \firstname{Marten Ole} \lastname{Schmidt}\inst{4}\fnsep\thanks{\email{ole.schmidt@cern.ch}} \and
        \firstname{Ruben} \lastname{Shahoyan}\inst{1}\fnsep\thanks{\email{ruben.shahoyan@cern.ch}} for the ALICE Collaboration
}

\institute{European Organization for Nuclear Research (CERN), Geneva, Switzerland
\and
           Frankfurt Institute for Advanced Studies, Ruth-Moufang-Str.~1, 60638 Frankfurt, Germany
\and
           Goethe University Frankfurt, Germany
\and
           University of Heidelberg, Germany
          }

\abstract{
  In LHC Run~3, ALICE will increase the data taking rate significantly to continuous readout of 50 kHz minimum bias Pb$-$Pb collisions.
  The reconstruction strategy of the online offline computing upgrade foresees a first synchronous online reconstruction stage during data taking enabling detector calibration, and a posterior calibrated asynchronous reconstruction stage.
  We present a tracking algorithm for the Time Projection Chamber (TPC), the main tracking detector of ALICE.
  The reconstruction must yield results comparable to current offline reconstruction and meet the time constraints like in the current High Level Trigger (HLT), processing 50 times as many collisions per second as today.
  It is derived from the current online tracking in the HLT, which is based on a Cellular automaton and the Kalman filter, and we integrate missing features from offline tracking for improved resolution.
  The continuous TPC readout and overlapping collisions pose new challenges: conversion to spatial coordinates and the application of time- and location dependent calibration must happen in between of track seeding and track fitting while the TPC occupancy increases five-fold.
  The huge data volume requires a data reduction factor of 20, which imposes additional requirements: the momentum range must be extended to identify low-$p_{\rm T}$ looping tracks and a special refit in uncalibrated coordinates improves the track model entropy encoding.
  Our TPC track finding leverages the potential of hardware accelerators via the OpenCL and CUDA APIs in a shared source code for CPUs, GPUs, and both reconstruction stages.
  Porting more reconstruction steps like the remainder of the TPC reconstruction and tracking for other detectors will shift the computing balance from traditional processors to GPUs.
  We give an overview of the foreseen tracking in Run~3 and discuss the track finding efficiency, resolution, treatment of continuous readout data, and performance on processors and GPUs.
}

\maketitle

\section{Introduction}
\label{intro}

ALICE (A Large Ion Collider Experiment~\cite{bib:alice}) is one of the four large experiments at the LHC (Large Hadron Collider) at CERN.
As a dedicated heavy-ion experiment, \mbox{ALICE} is primarily interested in Pb$-$Pb (lead) collisions.
During the long shutdown 2 (LS2) until 2021, the LHC upgrades will boost the Pb$-$Pb collision rate from currently around 10\,kHz to 50\,kHz.
With its current drift detectors in triggered readout mode and with the TPC gating grid, ALICE is limited to recording events at a rate in the order of~1\,kHz.
Usually, in high energy physics fast trigger detectors evaluate all collisions based on a small amount of fast measurements.
They select the physics-wise interesting events and trigger the readout of the remaining detectors.
However, the future ALICE physics program can be served best by storing all collisions without prior trigger selection~\cite{bib:aliceupgrade}.
In addition, due to out-of-bunch pile-up in the TPC (Time Projection Chamber), collisions recorded by ALICE in triggered mode always contain fractions of the previous and of the following collisions.
Consequently, at high interaction rates a large amount of data is unsuited for analyses since it belongs to collisions that are not fully recorded.
On top of that, this pile-up data complicates the event reconstruction.

\begin{figure}[htb]
\centering
\sidecaption
\includegraphics[width=8cm,clip]{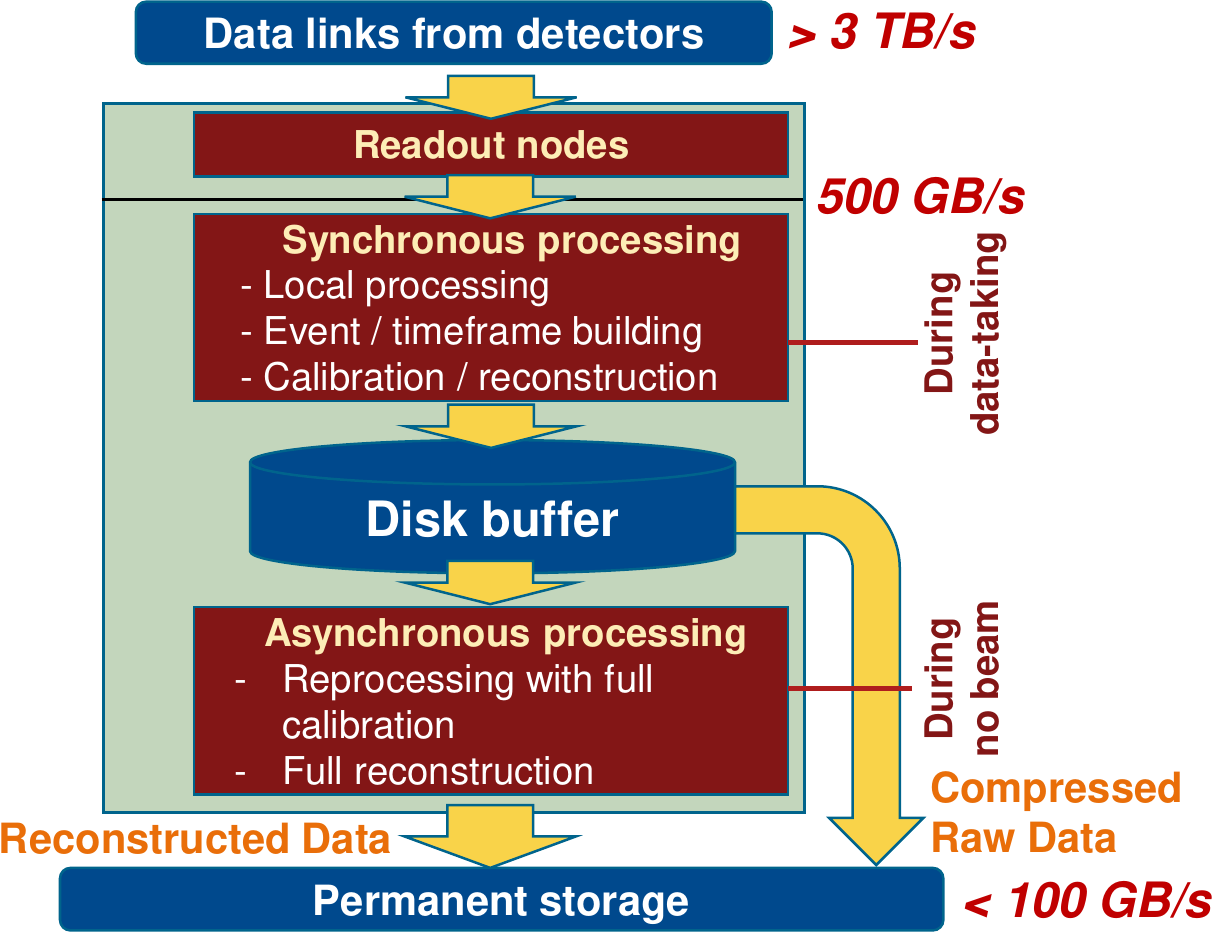}
\caption{Illustration of the ALICE O$^2$ computing scheme with two-phase event processing~\cite{bib:o2tdr}}
\label{fig:o2}
\end{figure}

For LHC Run~3 following the LS2, ALICE will thus change its readout strategy to a continuous one, i.\,e.~all minimum bias events will be fully recorded.
This goes along with a major upgrade of the main tracking detector TPC~\cite{bib:tpcrun3tdr} and with the ALICE O$^2$ project (Online Offline Computing Upgrade~\cite{bib:o2tdr}).
The O$^2$ computing scheme brings about a change of paradigm:
Instead of two independent reconstruction infrastructures (online, which is fast but very limited, and offline, which is precise but slow) a common software framework will be put in place.
There will be only one reconstruction algorithm, which may run twice with different calibrations and configurations.
Most reconstruction will run on the online compute farm at the experiment.
A first processing step synchronously to data taking performs data compression and the reconstruction steps required for the real-time calibration.
The output is stored onto a disk buffer.
Postprocessing steps can finalize the calibration thereafter.
When there is no stable beam in the LHC (during refills, technical stops, machine development, or shutdown at the end of the year), the available compute resources are used for an asynchronous full reprocessing of the data with the final calibration.
Figure~\ref{fig:o2} illustrates the described strategy.

These upgrades pose large challenges for the reconstruction and the computing.
Not only does the event processing rate increase fifty-fold from about~1\,kHz to~50\,kHz, but also the amount of data to be stored increases proportionally.
(The factor for the storage is less than~50, since the triggered events of Run~2 contains parts of additional pile-up events, while the continuous data in Run~3 will contain exactly the 50\,kHz of minimum bias events.)
It is illusionary to assume that compute or storage capacity will increase by more than an order of magnitude until 2021, and neither will a larger IT budget be able to compensate this.
Therefore new fast algorithms are inevitable to significantly speed up the event reconstruction and data compression techniques must reduce the data to a manageable size.
Additional challenges for the reconstruction stem from the new GEM TPC readout (Gas Electron Multiplier)~\cite{bib:tpcrun3tdr} and the new conditions:
The higher particle occupancy in the TPC does not only increase the combinatorial complexity, but in particular in combination with the ion back-flow from the GEM readout leads to significant space charge distortions inside the TPC, which must be corrected.
Furthermore, the continuous readout lacks the assignment of hits in the TPC to primary vertices, which prevents the direct computation of absolute $z$-coordinates in the TPC (the $z$-axis is oriented along the TPC drift direction).
Finally, the absence of a separate Offline reconstruction stage requires that online processing yields offline quality results at online processing rates.

\begin{figure}[htb]
\centering
\sidecaption
\includegraphics[width=8cm,clip]{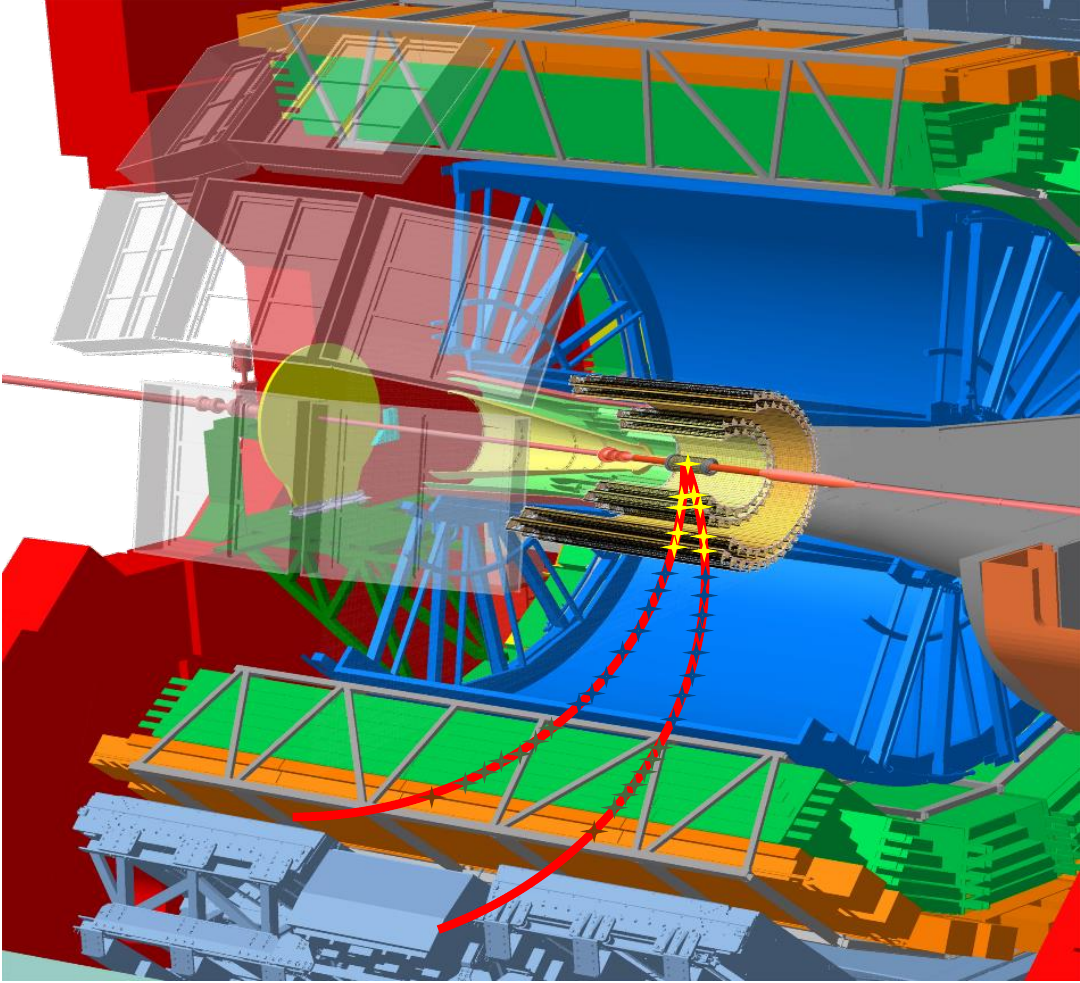}
\caption{Illustration of ALICE highlighting the detectors relevant for track reconstruction: yellow: 7 layers of silicon pixels in ITS (Inner Tracking System), blue: TPC (Time Projection Chamber) with 152 pad rows, green: 6 layers in TRD (Transition Radiation Detector), and orange: 1 TOF hit (Time Of Flight)}
\label{fig:alice}
\end{figure}

ALICE has three main tracking detectors, which include the ITS (Inner Tracking System)~\cite{bib:itsrun3tdr}, TPC (Time Projection Chamber)~\cite{bib:tpcrun3tdr}, and TRD (Transition Radiation Detector)~\cite{bib:trdnim}.
Optionally, a hit in TOF (Time Of Flight) can be used.
Figure~\ref{fig:alice} shows the geometry.

This paper discusses the ongoing improvements to the TPC tracking in detail.
We give a brief overview of the algorithm and the implementation and illustrate the steps needed to improve the track resolution obtained online to an offline level.
Next, we discuss the modifications with respect to the continuous TPC read out.
We give an overview of the TPC compression and a global picture including the other tracking detectors of ALICE.
Finally, we discuss the processing speed we achieve and give an outlook to the next steps.

\section{TPC Tracking Algorithm and Implementation}
\label{sec:algo}

The ALICE O$^2$ tracking algorithm for the TPC is derived from the existing HLT (High Level Trigger) TPC tracking, which starts with a Cellular Automaton based seeding followed by Kalman Filter and track following~\cite{bib:tns,bib:chep,bib:cnna}.
In a nutshell, the algorithm splits the volume of the TPC in 36 trapezoidal sectors.
The first phase finds track segments inside the sectors.
It starts with a Cellular Automaton that identifies links between hit triplets arranged close to a straight line, followed by an evolution step that concatenates consecutive links.
This yields short track segments of usually 4 to 10 hits, which are fit with a simplified Kalman Filter.
The fitted particle trajectory is extrapolated through the TPC volume, adjacent hits are collected without following multiple track hypotheses, and the fit is refined.
The second phase merges the track segments found in the individual sectors and performs the final track fit with the full Kalman filter.

In order to leverage modern architectures, the algorithm is implemented in a parallel way supporting multi-threading on the processor via OpenMP and offloading to GPUs via the CUDA and OpenCL APIs.
The majority of the code is kept in a common language for all these backends to avoid the overhead for maintaining multiple versions of the code~\cite{bib:generic}.

The algorithm was originally developed with hard timing constraints in mind to be deployed in the ALICE HLT.
It features a competitive tracking efficiency as compared to Offline tracking at the cost of a slightly worse resolution~\cite{bib:lhcp2017}.
In some aspects the online HLT tracking benefits from its different design, e.\,g.~the Cellular Automaton without strict vertex constraint yields better tracking efficiency for secondaries as compared to the Offline tracking, and its approach with sector-local tracking and track merging achieves a lower clone-rate.
In the next section, we illustrate how we have improved upon these results, to match the Offline tracker also in terms of momentum and spatial resolution.

The parallelization is implemented over the TPC hits during the Cellular Automaton steps and over the TPC tracks for the Kalman Filter and track following.
Despite the vast amount of hits in central Pb$-$Pb collisions, the number of tracks has become insufficient to fully load modern GPUs.
We account for this in the HLT by processing two events concurrently on one GPU~\cite{bib:chep2016gpu}.
In the O$^2$ computing, the processing of continuous data will be based on time frames of~10 to~23\,ms.
These time frames will contain many more hits and tracks, and we will be able to make full use of the GPUs by processing full time frames or large fractions thereof at once.

\section{Improvements of Tracking Resolution}

The ALICE HLT TPC tracking features identical or better efficiency compared to Offline but worse resolution.
Several causes have been identified leading to the deficiencies shown in~\cite{bib:lhcp2017}, and features from Offline tracking have been ported to improve the resolution.
\begin{compactitem}
\item The HLT tracking did not consider the $B_{\text{x}}$ and $B_{\text{y}}$ components of the magnetic field, which led to a systematic bias in the reconstructed $p_{\text{T}}$ depending on the pseudorapidity~$\eta$.
  The Track propagation has thus been adjusted to consider the full field, and the polynomial approximation of the field has been extended to three dimensions.
\item The HLT tracking is now employing the same cluster error parameterization as the Offline tracking.
\item The HLT tracking has been considering the initial seed from the Cellular Automaton as ground truth of the track, which is extrapolated first outward and then inward.
  This has been changed such that during the second extrapolation inward, the track is also propagated through the initial seed and outlier clusters are rejected.
\item The rejection of outlier clusters during the final track fit depending on the predicted~$\chi^2$ has been implemented.
\item The cluster error is enlarged for the following type of special clusters: clusters with two charge peaks split into two, clusters spanning only a single TPC pad row or only a single time bin, clusters at the edge of a TPC pad row where a fraction of the charge was not recorded, and clusters shared between multiple tracks.
\item The final Kalman refit is iterated three times: inward, outward, and inward, for a better cluster rejection and a better linearization of the non-linear track model.
\end{compactitem}

\begin{figure}[t!]
\centering
\includegraphics[width=12.1cm,clip]{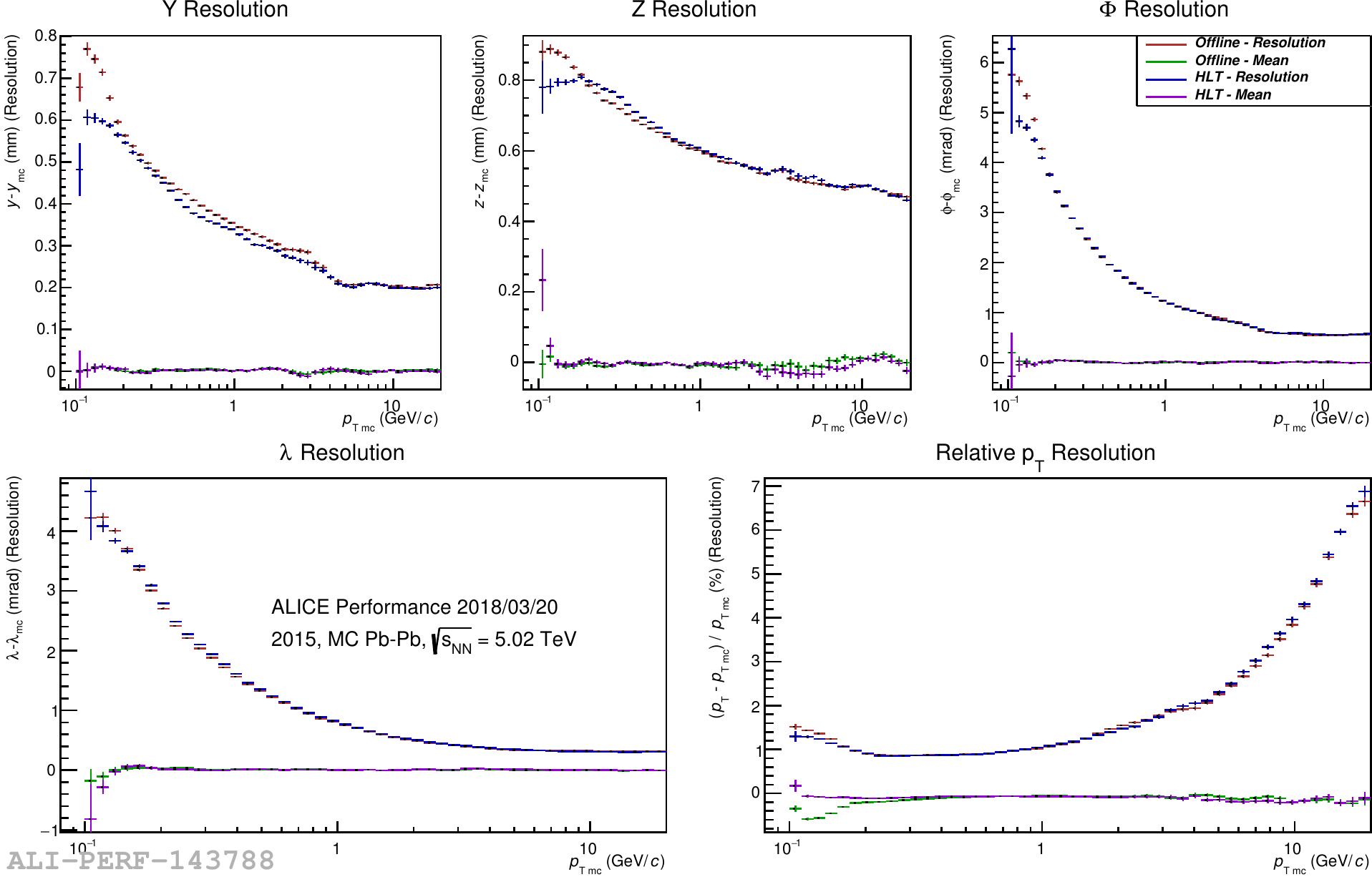}
\caption{Comparison of HLT/O$^2$ and Offline track resolution in Run~2 Pb$-$Pb without TPC distortions$^1$}
\label{fig:hltofflinenodist}
\end{figure}

The changes increase the compute time of the track fit significantly (currently by the factor three, but the new code is not yet optimized for performance), but since the majority of the time is spent for track finding, the total performance loss is acceptable.
We note that the new version also has a better low-$p_{\text{T}}$ track finding efficiency (see next section) and the Kalman fit of more tracks consequently takes more time.
\stepcounter{footnote}\footnotetext{Offline and HLT refer to the Offline and the HLT/O$^2$ versions of tracking, Resolution and Mean refer to the width and the mean (bias) of a Gaussian fit to reconstructed and Monte Carlo track parameters in slices of transverse momentum.}
A comparison of HLT and Offline resolution is shown in Fig.~\ref{fig:hltofflinenodist}.$^1$
The figure compares the resolution and the bias (mean) of the five track parameters $x$, $y$, $\phi$, the dip angle $\lambda$, and $p_{\text{T}}$ using the HLT and offline tracking.
The resolution of the HLT tracks has improved to Offline level with only marginal differences of at most~5\% in spatial resolution and~3\% in momentum resolution.

\begin{figure}[t!]
\centering
\includegraphics[width=12.1cm,clip]{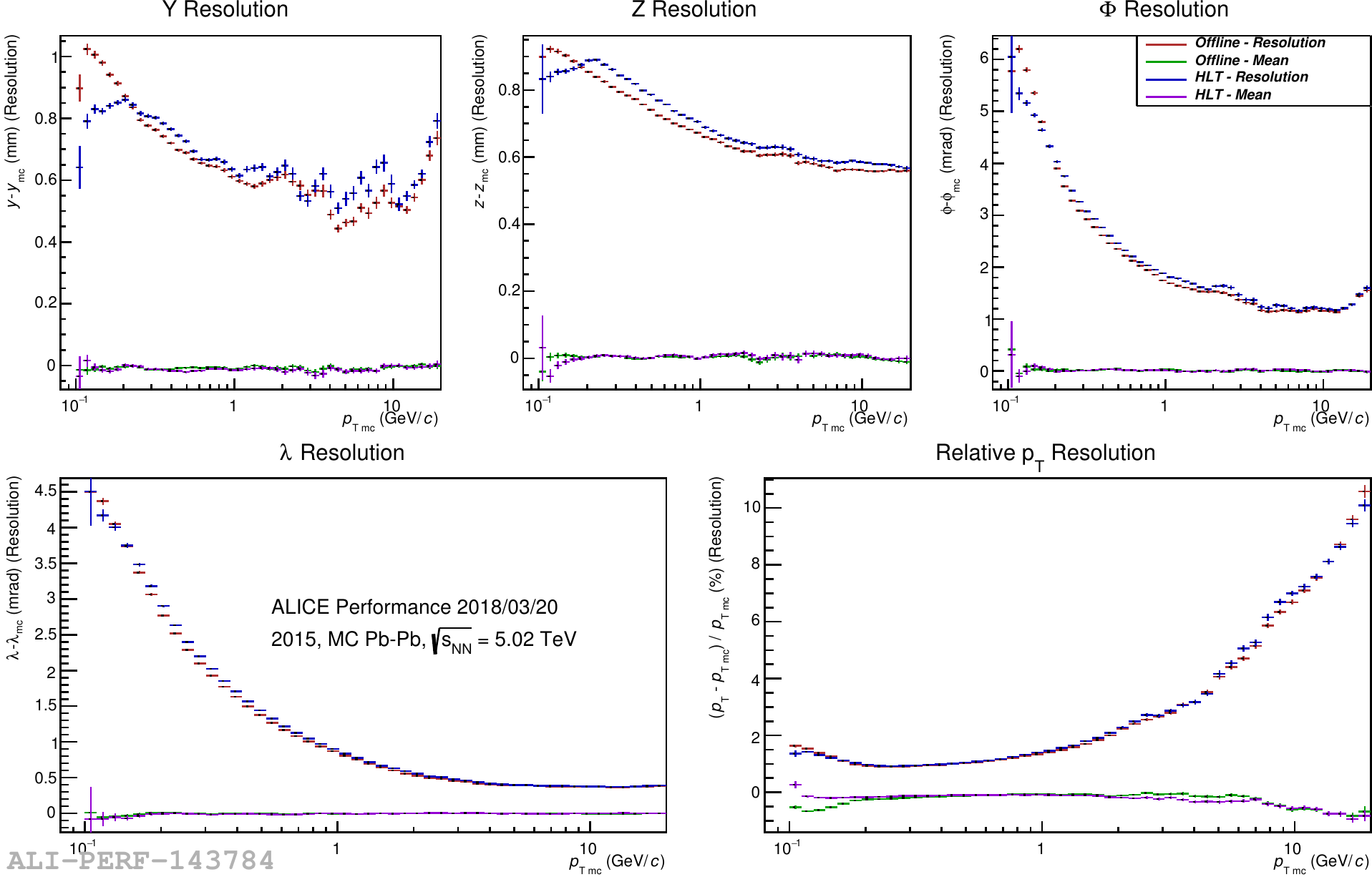}
\caption{Comparison of HLT/O$^2$ and Offline track resolution in Run~2 Pb$-$Pb with TPC distortions}
\label{fig:hltofflinedist}
\end{figure}

Figure~\ref{fig:hltofflinenodist} is based on TPC-only tracking without ITS refit, and uses Monte Carlo simulations of Pb$-$Pb data without space charge distortions.
For comparing the tracking algorithms, both HLT and Offline algorithms were running with identical cluster error parameterization and calibration (while the final calibration is not available on the HLT cluster).
The same figure for pp data shows a similar behavior with identical HLT and Offline results.
Offline is not generally better, but in particular for low momentum the HLT tracking features the better spatial resolution.
Even though much smaller than expected for Run~3, also Run~2 data is affected by TPC space charge distortions at high interaction rates.
Figure~\ref{fig:hltofflinedist} shows the same quantities as Fig.~\ref{fig:hltofflinenodist} with TPC distortions in the simulation.
With distorted data, the resolutions are generally 5\% to 30\% worse than without distortions, but the behavior of HLT and offline tracking is basically identical.
The Run~2 Offline tracking already contains special treatment, e.\,g. the covariance is adjusted for the correlation of the systematic cluster errors added in highly distorted regions.
In contrast, the HLT tracking does not yet have treatment for distortions, except for the usage of the same increased statistical errors.
This leads to slightly worse spatial resolutions with HLT tracking in comparison to Offline tracking.

\section{Tracking of Continuous Data in Time Frames}
\label{sec:trackingtf}

\begin{figure}[htb]
\centering
\includegraphics[width=12.5cm,clip]{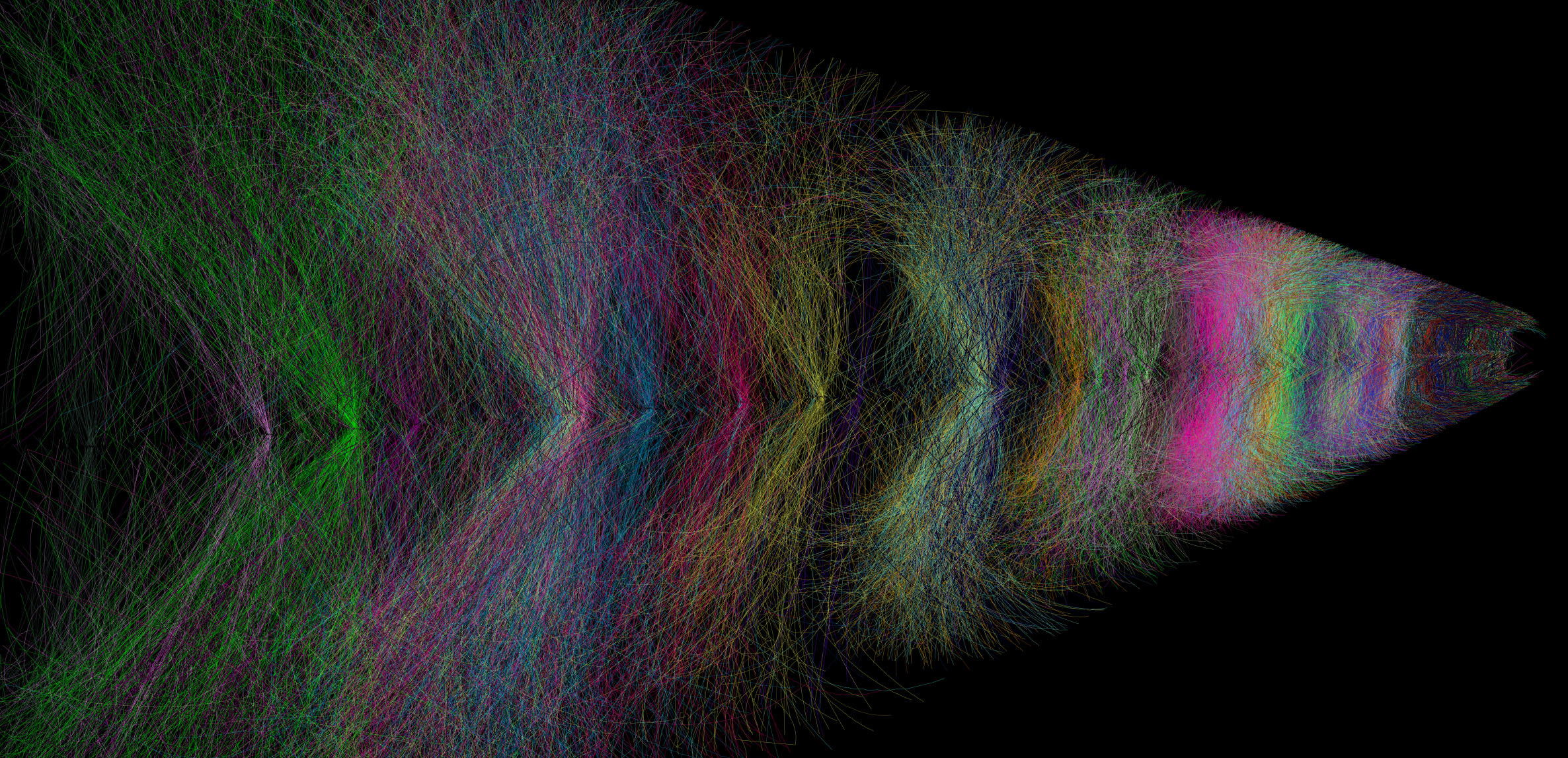}
\caption{Cross section of a time frame as expected from the TPC in Run~3 with time time scaled linearly to $z$ and tracks from different primary collisions shown in different color}
\label{fig:timeframe}
\end{figure}

The HLT tracking has been extracted from the Run~2 software into a dedicated common standalone package that is compatible with HLT software for Run~2 and O$^2$ software for Run~3.
Different header files are used for certain constants (e.\,g.~related to the geometry).
As a first step, the TPC cluster time is scaled linearly to compute the $z$-coordinate, and the tracking runs assuming an arbitrarily long TPC.
Therefore, the events are shifted in $z$ by a factor proportional to the collision time, as shown in Fig.~\ref{fig:timeframe}.
Eventually, the tracking shall work with native TPC pad, row, and time coordinates directly to cope with the distortions.
This still needs to be implemented, and will in addition require to perform the transformation of native into spatial coordinates on the fly during the tracking on the GPU.
Having reached similar performance in Run~2 online tracking, the tracking needs to cope with the particularities of the continuous read out.
The most important aspects are:
\begin{compactitem}
\item There is no a priory knowledge which TPC hit belongs to which primary vertex.
  Consequently, the TPC clusters have no defined $z$-coordinate.
\item The seeding assumes no vertex constraint due to the uncertainty of the $z$-position.
\item The TPC space point calibration correction, the inhomogeneous magnetic field, and the cluster error parameterization depend on the $z$-position of the hit, and can thus only be applied after assigning the hit to a primary collision vertex.
\end{compactitem}

\begin{figure}[b!]
\centering
\includegraphics[width=12.1cm,clip]{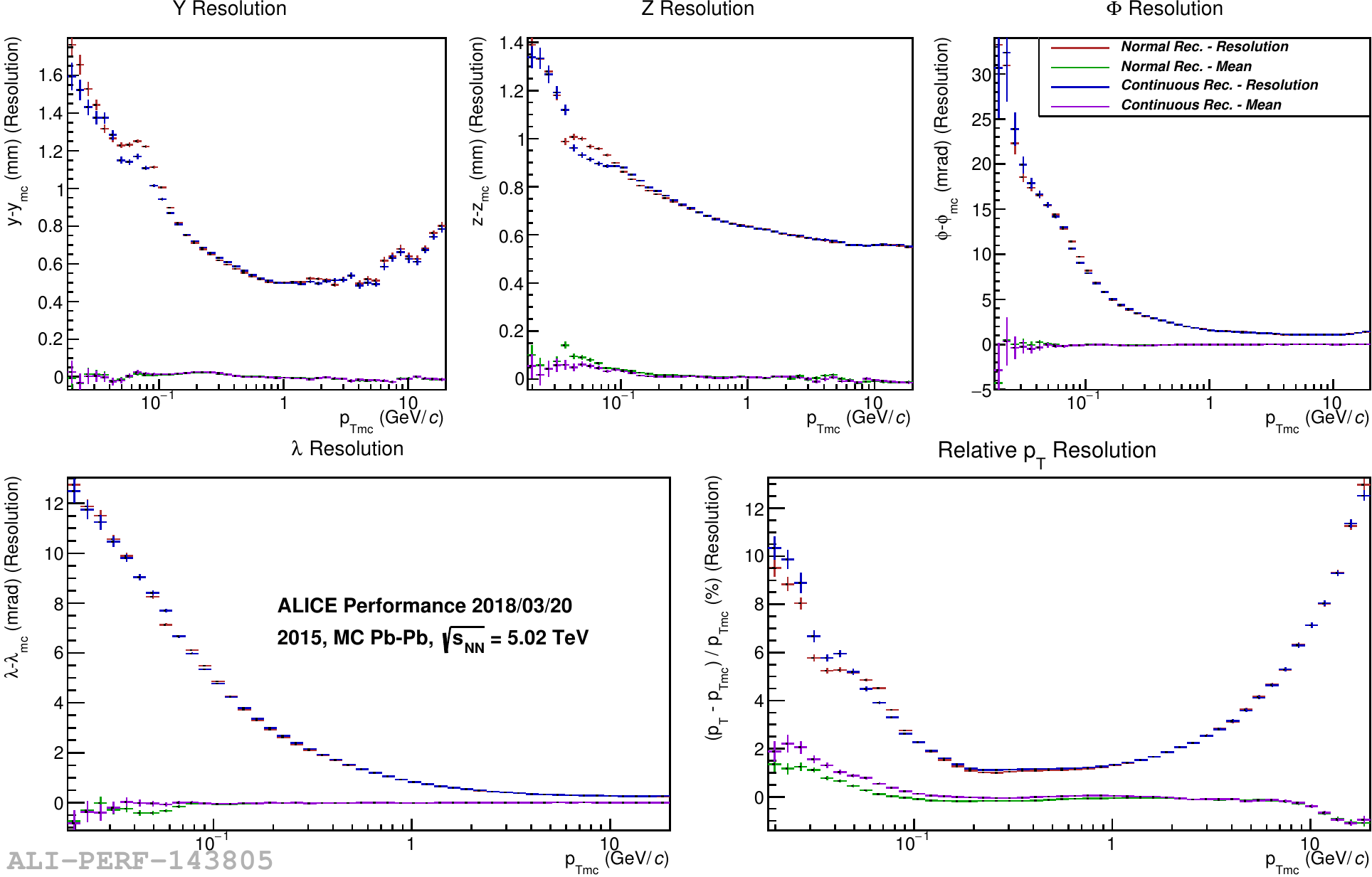}
\caption{Comparison of O$^2$ tracking resolution between normal tracking of single events and $z$-independent continuous tracking of time frames}
\label{fig:o2z}
\end{figure}

The TPC tracking for O$^2$ treats this in the following way:
\begin{compactitem}
\item The TPC space charge distortions are smooth, therefore the Cellular Automaton seeding can still find the track segments.
\item The seeding uses a search window in~$\eta$ instead of a vertex window, eliminating the loose vertex constraint.
\item After the Kalman fit of the initial seed, the position of closest approach to the beam line is computed and assumed as $z$-position (and thus time) of the primary vertex.
  By design, this estimate is wrong for deep secondaries from long-lived decays.
\item Later, this could be refined by matching with the known times of the primary vertices.
\item This yields a preliminary estimate of the absolute $z$-position, such that corrections, cluster errors, and the magnetic field can be computed.
\item The track is propagated through the TPC volume and additional clusters are picked up.
\item The individual track segments are merged (see section~\ref{sec:algo}), the vertex estimate is refined with the improved track fit, and the track is shifted in $z$ to its new position.
\item In case clusters of a fully merged secondary track are outside the real TPC volume, the track is shifted inside, partially fixing the incorrect vertex estimate for secondaries.
\item After the first inward propagation of the track fit, the vertex estimate is again refined.
\item At all times, track parameters are kept at coordinates inside the real TPC volume to avoid single precision floating point artifacts, which occur in the track fit at large $z$.
\end{compactitem}

In addition to the time frame support, much effort was put into tracking of low-$p_{\text{T}}$ particles, which is needed for data compression (see section~\ref{sec:comp}).
This benefits from the modified Cellular Automaton with~$\eta$-search-window, since additional legs of looping tracks do not point to the vertex.
In addition, the refit was tuned to enable arbitrary rotations for following the full helix, and several cuts were modified.
Still ongoing is the implementation of a special low-$p_{\text{T}}$ track merging since the current merging procedure does not cover all cases.

In the following, we evaluate the track resolution and efficiency with the above-mentioned changes.
First, Fig.~\ref{fig:o2z} compares the resolution of the traditional tracking of single events with the tracking of continuous data by forgetting the absolute $z$-information of the triggered events and processing them as a time frame.
It demonstrates that the track resolution is practically identical, except for small differences at very low momentum, where the purpose of the tracking lies anyway mostly in data compression.
Further there is no difference in the track finding efficiency.

\stepcounter{footnote}\footnotetext{Efficiency: fraction of reconstructed MC tracks (with at least 1 MC hit in the TPC), Clone rate: fraction of tracks reconstructed multiple times v.s. all tracks, Frake tracks: fraction of tracks with more than~10\% fake cluster attachmend v.s. all tracks, Findable: At least 70 MC hits in the TPC.}

\begin{figure}[t!]
\centering
\includegraphics[width=12.1cm,clip]{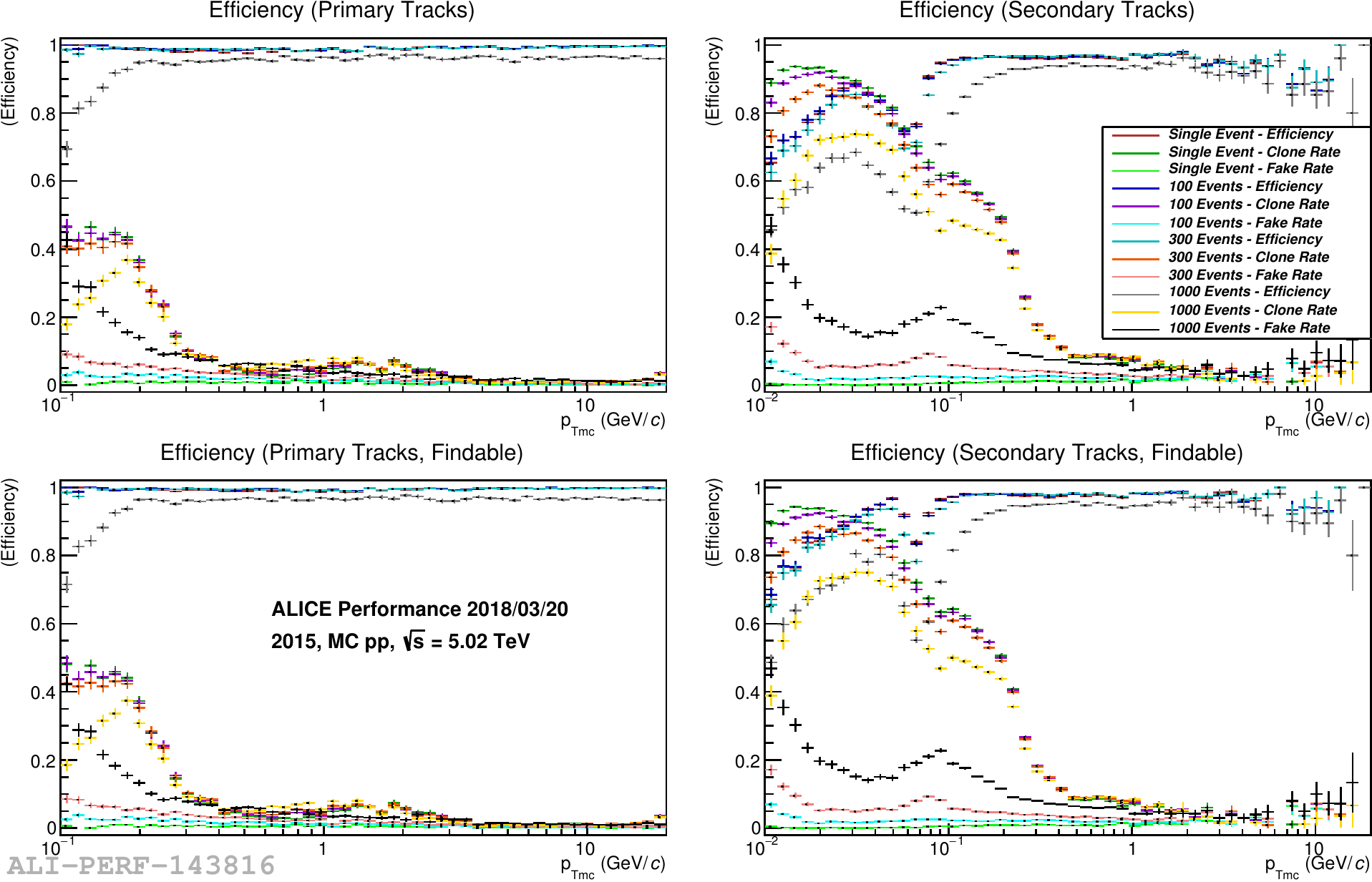}
\caption{Comparison of the O$^2$ tracking efficiency of pp data for different pile-up scenarios$^2$}
\label{fig:o2pileup}
\end{figure}

Next, we compare the tracking at different TPC occupancies.
We start with overlaying  pp collisions.
Since we run only TPC tracking without ITS, we do not consider vertexing at all.
Figure~\ref{fig:o2pileup} compares the efficiency of single events with a pile-up of up to~$\mu = 1000$.$^2$
While up to~$\mu = 100$ there is absolutely no effect, a pile-up of~$\mu \geq 300$ results in a small decrease in the secondary finding efficiency at low~$p_{\text{T}}$.
The extreme scenario of~$\mu = 1000$ finally shows a general significant reduction of efficiency, and large increase of fake rate.
We note that 50\,kHz Pb$-$Pb is equivalent to~$\mu = 450$ in pp.

\begin{figure}[t!]
\centering
\includegraphics[width=12.1cm,clip]{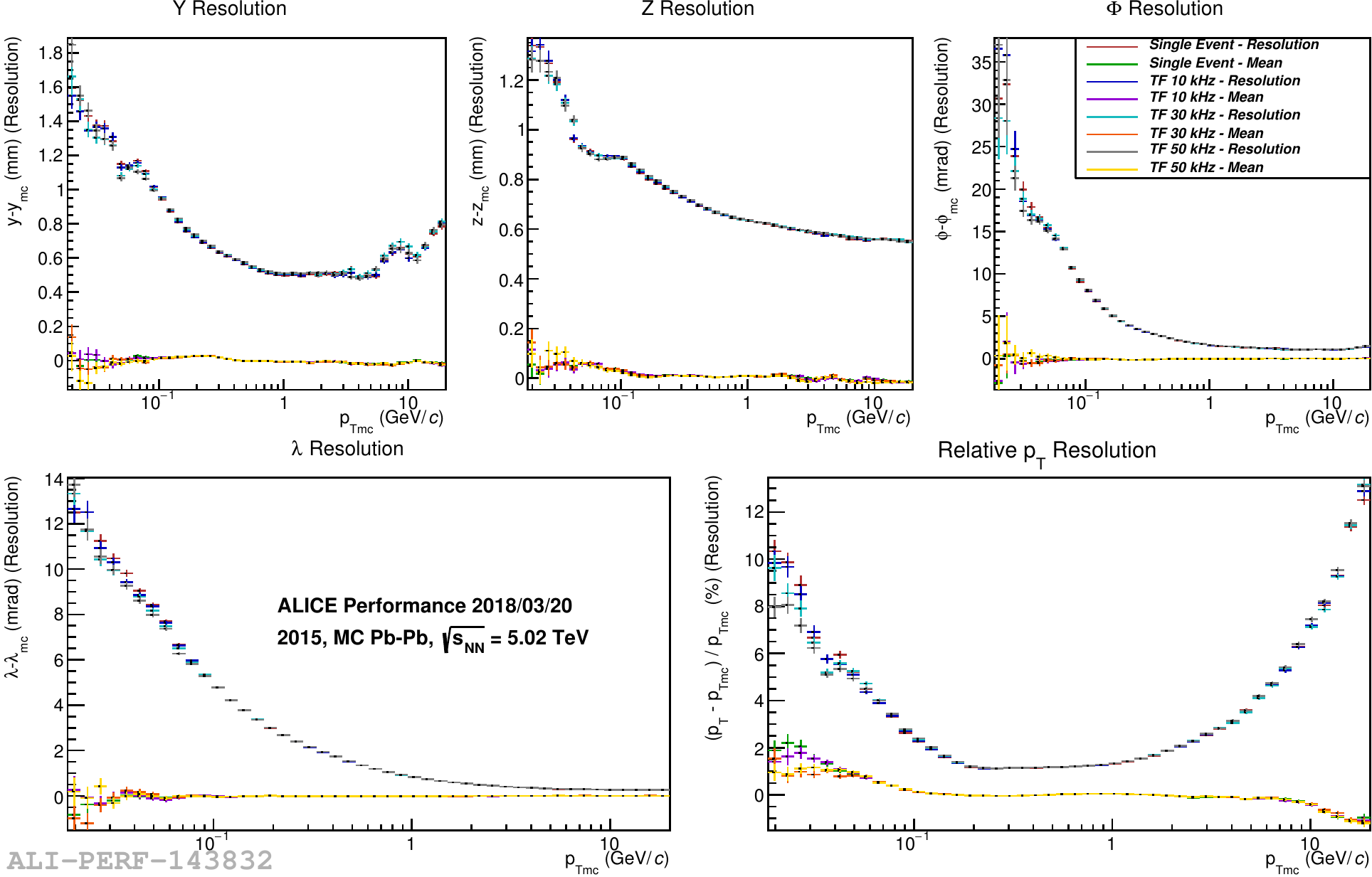}
\caption{Comparison of the O$^2$ tracking resolution obtained using different interaction rates}
\label{fig:o2res}
\end{figure}

\begin{figure}[t!]
\centering
\includegraphics[width=12.1cm,clip]{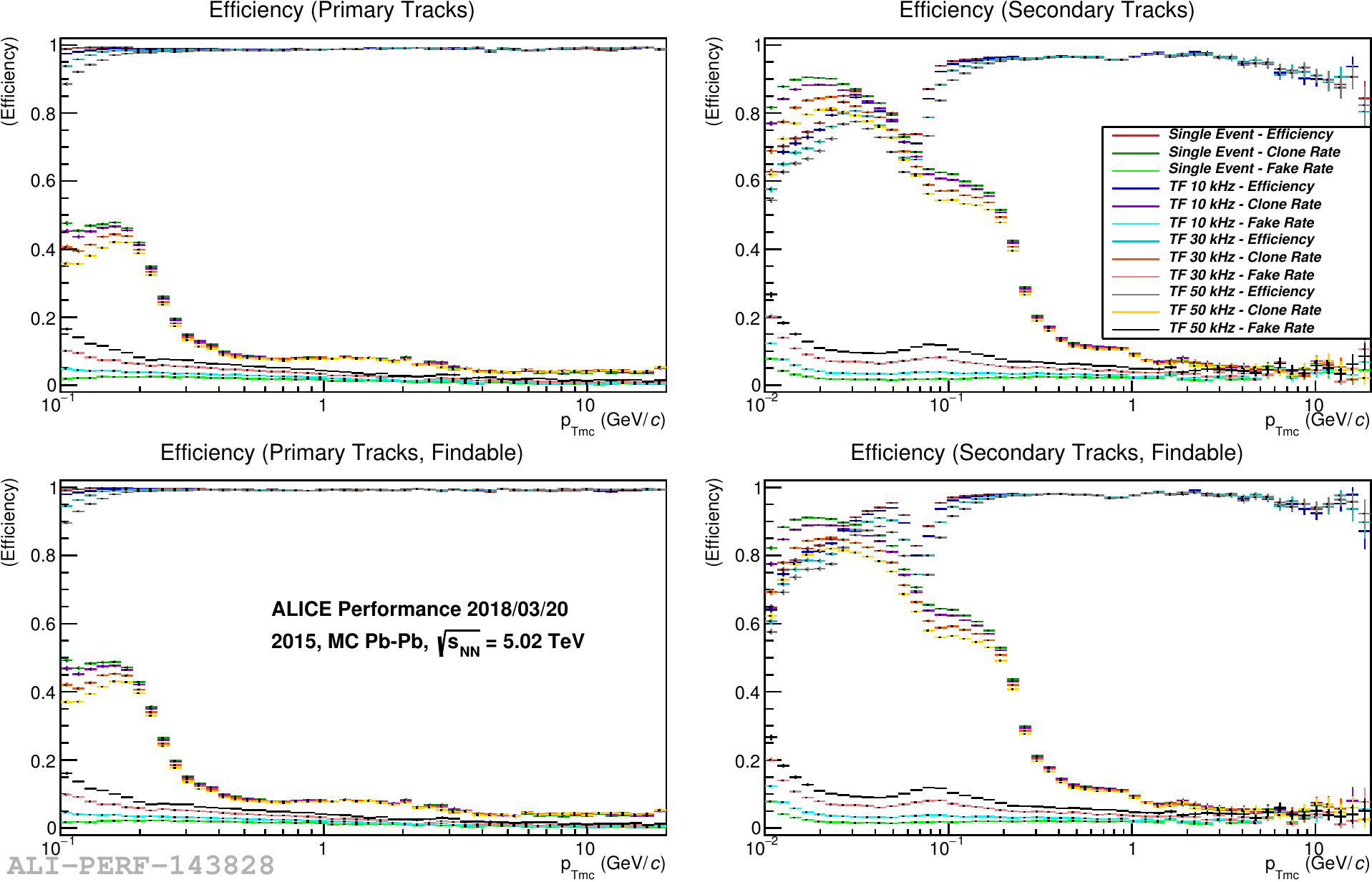}
\caption{Comparison of the O$^2$ tracking efficiency obtained using different interaction rates}
\label{fig:o2eff}
\end{figure}

Finally, we compare the continuous O$^2$ tracking at different interaction rates to single events in a time frame.
This uses a realistic bunch crossing simulation with a Poisson distribution.
Figures~\ref{fig:o2res} and~\ref{fig:o2eff} show the resulting tracking resolution and efficiency.
While the resolution does not suffer from higher occupancy at all, there are some small differences in the efficiency.
Starting from 30\,kHz, the track finding efficiency decreases below 200\,MeV/$c$ in~$p_{\text{T}}$.
The fake rate increases with higher occupancy as expected.
We want to underline the high efficiency for low-$p_{\text{T}}$ tracks of around~70\% down to~15\,MeV/$c$.
The high clone rate is a consequence of the incomplete low-$p_{\text{T}}$ track merging and reflects that we find many legs of looping tracks.

\section{Global Reconstruction}

This section gives an overview of reconstruction tasks related to the TPC tracking, primarily TPC data compression which is based on the TPC tracking, and the tracking of the other detectors.

\subsection{Data Compression}
\label{sec:comp}

\begin{figure}[htb]
\centering
\sidecaption
\includegraphics[width=6cm,clip]{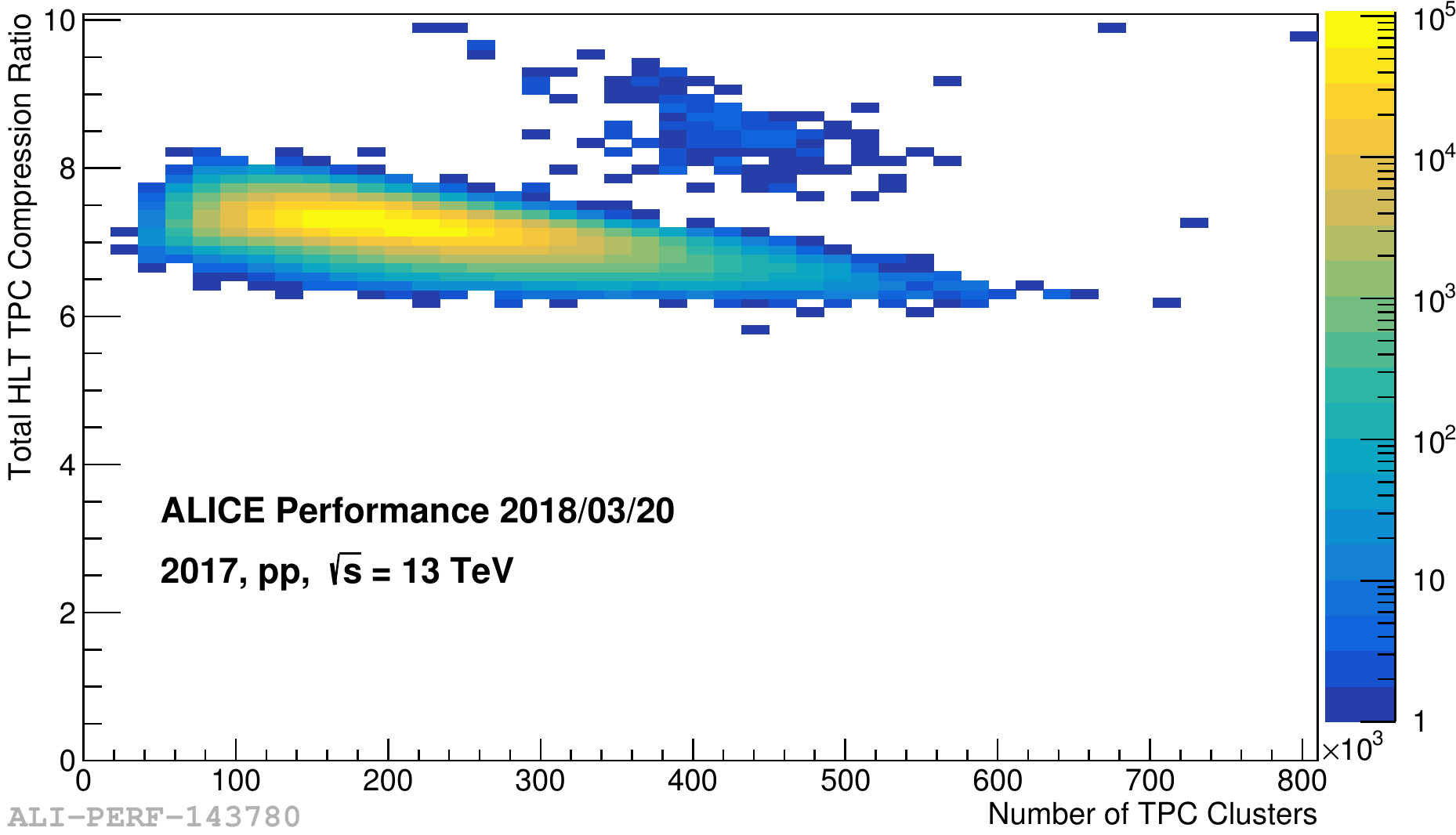}
\caption{Total ALICE TPC data compression factor v.s. number of clusters for pp events recorded in 2017}
\label{fig:compression}
\end{figure}

In order to fit the number of events to be recorded in Run~3 into the allocated disk space, the TPC data size must be compressed by a factor 20 (compared to Run~2 raw data size\footnote{The Run~3 raw data format will be different but this is irrelevant since ALICE stores only TPC clusters reconstructed online. Therefore, we fix the current raw size as the baseline for comparisons.}).
The data compression for O$^2$ will consist of four steps: cluster finding, rejection of clusters not relevant for physics, entropy reduction, and entropy encoding.
All steps except for the cluster rejection are already implemented for Run~2, achieving a TPC compression factor of 7.2 for 2017 pp data (see Fig.~\ref{fig:compression}).
The largest compression factor comes from the actual entropy encoding, but the cluster finding and entropy reduction steps are needed to enable efficient entropy encoding.
For comparison, the Huffman entropy encoding of the TPC ADC values directly achieves only a compression factor of up to 2.

The Run~3 entropy reduction will improve upon Run~2 by employing a track model compression, i.\,e.~storing only the residuals for clusters attached to tracks.
The entropy of these residuals is decreased further by refitting the track in a distorted coordinate system to avoid calibration effects~\cite{bib:lhcp2017}.
The prototype of the Run~3 compression improves the compression factor from 7.2 to 9.1, compared to Run~2.
We aim to obtain the missing factor of around 2 by removing clusters not used for physics.
This includes very low-$p_{\text{T}}$ looping tracks below~50\,MeV/$c$, additional legs of tracks below~200\,MeV/$c$, and track segments with high inclination angle.
From the tracking efficiency, we assume we can identify the clusters down to~10$-$15\,MeV/$c$ during tracking, where the track finding efficiency is~70\% or higher (see Fig.~\ref{fig:o2eff}).
Other methods are being investigated for the $p_{\text{T}}$ regime below (e.\,g. Hough Transformation, Neural Networks).

\subsection{TPC Calibration}

As explained in section~\ref{sec:trackingtf}, the ions drifting from the GEM readout to the central electrode create significant space charge distortions in the TPC.
The calibration procedure involves a an ITS-TRD-only refit of global ITS-TPC-TRD tracks.
The corrections can then be obtained from the residuals of the TPC hits~\cite{bib:lhcp2017}.
This procedure requires online tracking of ITS and TRD.
In contrast to the TPC online tracking which must process all events for the compression, ITS and TRD tracking of a subset of the events is sufficient if they contain enough tracks for the calibration.
In addition, it is not relevant to achieve close to 100\% tracking efficiency.

Beside this full TPC calibration available in the asynchronous phase, we consider a partial online calibration to correct for the majority of the average distortions.
This can be implemented via a feedback loop, as it is currently used in the ALICE HLT for the TPC drift velocity calibration in Run~2~\cite{bib:chep15, bib:tns2016}.
The scheme requires the usage of flat data objects for reconstruction and calibration avoiding serialization and deserialization overhead.
This has been prototyped in the HLT~\cite{bib:ctd} and will be handled similarly in the O$^2$ software.

\subsection{ITS Tracking and Matching}

In Run~2, ALICE is relying on the TPC to ITS prolongation, which is much faster than standalone ITS tracking given the high occupancy in ITS.
Due to distortions in the TPC, this will not work in Run~3.
Consequently, ALICE is developing a standalone ITS tracking that will also be able to leverage hardware acceleration via GPUs.
A fast version will run in the synchronous phase, targeting only tracks which can be identified easily, e.\,g.~without missing hits in the ITS, which is sufficient for the calibration step.
The full ITS tracking will run in the asynchronous phase for the final reconstruction.
Thereafter, the TPC and ITS standalone tracks are merged.
It is conceivable to run TPC to ITS prolongation tracking on top with the final TPC calibration to attach remaining ITS hits of tracks that were not reconstructed standalone.

\subsection{TPC-TRD Tracking}

Compared to the ITS, the TRD features more fake hits, which increases the combinatorics for standalone tracking even further.
Therefore, ALICE plans to rely on the TPC-TRD prolongation tracking also for Run~3.
A prototype of the TRD online tracking has been developed in the scope of the ALICE HLT and is running in production since 2018.
We are currently porting this tracker to the O$^2$ software and consider to run it on GPUs as well.

\subsection{Global Tracking}

The baseline is to have the full TPC tracking on GPUs as well as a part of ITS and TRD tracking.
Thinking a bit further, having the full chain of the TPC, ITS, and TRD tracking on the GPU opens the opportunity to run a significant fraction also of the asynchronous reconstruction steps on GPUs.
This would ensure good utilization of the GPUs also during the asynchronous reconstruction, where the contribution from TPC processing is much smaller than during the synchronous phase.
Ideally, the GPU would also process all related task needing the same data.
The TPC-ITS matching is well parallelizable and should profit from GPUs.
The same is true for the entropy reduction steps of the data compression, while it remains to be seen how ALICE will handle the final entropy encoding.
This would allow us to run also the final compute intense refit on the GPU, without ever transferring data between host and GPU.
We are evaluating which of these additional steps are feasible and valuable on top of the baseline scheme.

\section{Next Steps}

The next immediate steps are the integration of some missing features into the TPC tracking: primarily the full merging of low $p_{\text{T}}$ looping tracks and the merging and propagation of tracks crossing the TPC central electrode.
The most important milestones will then be to abandon the a priori transformation of TPC hits from native pad, row, and time coordinates to $x$, $y$, and $z$-coordinates.
Instead, the tracking will work in native coordinates and transform the hits on the fly once the time of the vertex has been determined.
On top, the full tracking will be benchmarked with data simulated with full realistic distortions.
This is currently being implemented into the O$^2$ software.
Once the proper merging of the TPC tracks is implemented, we will go on with the rejection of TPC hits not used for physics to achieve the necessary TPC compression factor of~20.
Porting of the HLT TRD tracking to the O$^2$ software must be finalized.
These steps should finalize the baseline tracking implementation.
On top of that, many performance optimizations are conceivable, in particular by bringing more reconstruction steps onto the GPU.
The silver bullet would be to have the full ITS, TPC, and TRD tracking and compression on the GPU without intermediate data transfer.

\section{Conclusions}

ALICE has developed a fast new tracking for the TPC in the O$^2$ software for Run~3 and 4, derived from the Run~2 HLT tracking.
It is based on Cellular Automaton and Kalman filter, and can run on CPUs and GPUs featuring a shared common source code.
We have added several features known from Run~2 Offline tracking to achieve the same track resolution as Offline, while the O$^2$ track finding efficiency is already equal to or better than the current Offline efficiency.
The new data taking scheme with continuous read out requires the tracking to work independently from absolute $z$-coordinates and without a primary vertex constraint.
The O$^2$ version has been adjusted accordingly, without significant decrease in efficiency or resolution.
Besides few technical features that still need to be implemented, the next important milestone and benchmark is the processing of fully distorted data in native TPC coordinates, performing the transformation to spatial coordinates on the fly.
On top of the baseline scheme with full TPC tracking and partial ITS and TRD tracking on the GPU, we aim to offload more reconstruction steps onto the GPU, in order to ensure the best possible usage of the GPU-accelerated online computing farm.
Newest benchmarks show a processing time of around 20 seconds for a 23\,ms time frame of 50\,kHz Pb$-$Pb data on an NVIDIA GTX 1080 GPUs.
This translates to a required GPU capacity of around 1000 cards for the TPC tracking, while we foresee 1500 servers with at least one GPU for the O$^2$ computing.
This leaves sufficient margin for the ITS and TRD tracking of a subset of events, online calibration, and other online processing tasks.

\end{document}